\documentclass[a4paper]{article}

\usepackage{amsmath,authblk,amssymb,graphicx,subfig,bm,lineno}
\usepackage[T1]{fontenc}
\usepackage[right=1in,left=1in,top=1in,bottom=1in,]{geometry}

\def\bold{\textbf}

\title{The effect of Confined One Gluon Exchange Potential and Instanton Induced Interaction on Nucleon-Nucleon Interaction}
\author[1,2]{S Raghavendra \footnote{ragavxyz@gmail.com}}
\author[3]{C Vanamali Shastry\footnote{vanamalishastry@gmail.com}}
\author[1,4]{V K Nilakanthan\footnote{nveluthat@gmail.com}}
\author[1]{K B Vijaya Kumar\footnote{kbvijayakumar57@gmail.com}}
\affil[1]{Department of Physics, Mangalore University, Mangalagangotri 574199, India}
\affil[2]{Department of Post-Graduate Studies in Physics, SDM College, Ujire 574240, India}
\affil[3]{Department of Education in Science and Mathematics, Regional Institute of Education Mysore, Mysuru 570006, India}
\affil[4]{Department of Post Graduate Studies in Physics, St. Aloysius College (Autonomous), Mangalore 575003, India}
\date{}

\begin{document}
\maketitle
\begin{abstract}
The effect of Confined One Gluon Exchange Potential and Instanton Induced Interaction potential in the singlet ($^1S_0$) and triplet ($^3S_1$) channels for Nucleon-Nucleon interaction has been investigated in the framework of Relativistic Harmonic Model using Resonating Group Method in the adiabatic limit with Born - Oppenheimer approximation. The contributions of the different components of the interaction potentials have been analyzed.
\end{abstract}
\bold{Keywords: }{Nucleon-Nucleon Interaction, Relativistic Harmonic Model, Confined One Gluon Exchange, Instanton Induced Interaction, Resonating Group Method.\\}
\bold{PACS:}{13.75.Cs, 12.39.Ki, 21.45.Bc}

\section{Introduction}

The problem of nucleon-nucleon (NN) interaction is one of the fundamental problems in physics. Ever since the discovery of the nucleus, the physics community has been striving to unravel the nature of the nuclear force. The study of NN interaction forms the foundation for this quest. With the advent of the quantum chromodynamics (QCD), the study of NN interaction took significant step, where in, it became necessary to include the mass spectra of the nucleons by taking in to account their quark structure.

In theory, almost any potential well with two adjustable parameters could be made to fit all low energy proton-proton ($p$-$p$) data, and a similar treatment for the neutron-proton ($n$-$p$) case holds as well \cite{Reid:1968sq}. De Rujula \textit{et al}., have explored the implications of this observation for the spectroscopy of the hadrons in the standard gauge model of weak, electromagnetic and strong interactions. The model \cite{DeRujula:1975qlm} involved four types of fractionally charged quarks, each in three colors, coupling to massless gauge gluons. The asymptotic freedom of the model was incorporated to argue that for the calculation of hadron masses, the short-range quark - quark interaction may be taken to be Coulomb-like. One can thus derive many successful quark-model mass relations for the (light) hadrons. The model gives a qualitative understanding of many features of the hadron mass spectrum, such as the origin and sign of the $\Sigma$-$\Lambda$ mass splitting. The model also predicts the masses of charmed mesons and baryons \cite{DeRujula:1975qlm,Warke:1980jq}.

Thus, NN interaction came to be explained by the exchange of various particles in the framework of QCD \cite{DeRujula:1975qlm,Machleidt:1987hj,Huang:2018rpb}. Since we neither know the exact form nor the nature of origin of confinement, various phenomenological quark models (both relativistic and non-relativistic) have been developed to explain NN interaction \cite{Warke:1980jq,Shimizu:1989ye,Oka:1981ri,DeTar:1977qf,Liberman:1977qs,Maltman:1983wx,Stancu:1997dq,Myhrer:1987af}.

Although much progress has been made in the understanding of the NN interaction, our understanding of the dynamics of the short-range interaction is still unsatisfactory. In the short range, NN interaction is repulsive in nature - a crucial attribute for nuclear stability. Typically, non-relativistic quark model (NRQM) Hamiltonians consist of kinetic energy, confinement potential and quark-quark interaction potential. Various potentials like one-gluon exchange potential (OGEP), instanton induced interactions (III), one-pion exchange potential (OPEP) and their combinations have been used to model the quark-quark interaction. In all these models, the short-range repulsion is attributed to the Pauli repulsion between the quarks manifesting itself as the exchange part of the color magnetic interaction \cite{Warke:1980jq,Isgur:1977ef}. The meson exchange models explain the short range repulsion through an interaction mediated by vector mesons \cite{Machleidt:1987hj}.

Although QCD is the accepted fundamental theory of strong interaction, there is no exact solution to the theory in the non-perturbative (low-momentum) regime. QCD has two stand-out features that set it apart from other interactions: color confinement and asymptotic freedom. Since the exact form of confinement is not known, we resort to \textit{phenomenological models}. In these models, the parameters in the theory are fitted to known hadronic properties such as masses and magnetic moments of the hadrons \cite{Isgur:1977ef,Chodos:1974pn}. The phenomenological models used in the past, like the relativistic MIT bag models \cite{Chodos:1974pn} and the non-relativistic potential models \cite{Warke:1980jq,Isgur:1977ef} as well as relativistic confinement models \cite{Khadkikar:1985pr}, have been successful in reproducing the hadronic spectra.

Oka \textit{et al.} have studied NN interaction by including the III along with OGEP. They show that, because III too has a color magnetic term, it can give rise to short range repulsions \cite{Oka:1989ud}. The inclusion of the III is justified because of the large value of the strong coupling constant ($\alpha_{s} \sim 1.6$) needed to reproduce the baryon spectrum in the phenomenological models that include only OGEP and the sizable contribution of the non-perturbative $q\bar q$ condensate to the N-$\Delta$ and $\pi$-$\rho$ splitting, as evident from the lattice QCD simulation in the quenched approximation \cite{Fukugita:1986tg}. Faessler \textit{et al} \cite{Faessler:1982ik} have analyzed the repulsive core using the quark model using the resonating group equation for NN scattering solved with the Breit-Fermi color interaction. They further used linear and quadratic potentials for the confinement. The color magnetic interaction adjusted to the nucleon mass splitting was shown to favor the orbital symmetry but not the completely symmetric orbital state and also addressed the infrared slavery problem.

Addressing the question of confinement, it is imperative to look at the confinement model for the gluons. Gluons being quanta of the color field, carry color charges and hence can interact among themselves, unlike the photons. Therefore, the effect of gluon exchange and confinement of gluons among relativistically confined quarks in NN interaction becomes indispensable \cite{Khadkikar:1991fv}. While QCD based models such as the MIT bag model \cite{DeTar:1977qf, Liberman:1977qs} or the non-relativistic-quark model (NRQM) \cite{Warke:1980jq,Oka:1981ri,Faessler:1982ik}, have incorporated the confinement of quarks using the resonating group method (RGM), the effect of confinement of gluons was unaccounted for. In these models, the color magnetic part of the Fermi-Breit one-gluon exchange potential (OGEP) is responsible for the short-range repulsion and $\sigma$ and $\pi$ mesons are used to obtain the bulk of NN attraction \cite{DeRujula:1975qlm}. These aspects were investigated from the perspective of color confinement of gluons by Vijaya Kumar \textit{et al.} \cite{Khadkikar:1991fv, VijayaKumar:1993np}. Therefore, to get a physically meaningful result, we have to incorporate the propagation of gluons in the confinement model \cite{Vinodkumar:1992wu}. This model was developed based on the study of the current confinement model (CCM) \cite{Khadkikar:1985pr,Khadkikar:1987kk} for the gluons and a Relativistic Harmonic Model (RHM) for the quarks \cite{Khadkikar:1991fv}. 

The instantons were introduced by `t Hooft in relation to the $U_{A}(1)$ problem. They play an important role in the effective interaction by coupling to the light quarks, whose strength of interaction depends on instanton density. This was estimated from the gluon condensate of the QCD vacuum \cite{Hooft:1976fv,Shifman:1979uw}. Moreover, lattice QCD suggests that the instantons density in the QCD vacuum is consistent with the gluon condensate expected from QCD sum rules \cite{Chu:1994vi}. In essence, the constituent quark masses is attributed to the spontaneous chiral symmetry breaking in QCD, and a plausible mechanism is due to instantons \cite{Shifman:1979uw,Shuryak:1984nq}. III has also been found necessary in explaining the spectra and decays of mesons \cite{Semay:1997ys,Bhavyashri:2005rb}. It is hence imperative that one includes III as a short range non-perturbative gluon effect \cite{Takeuchi:1995fq}.

The present paper attempts to reproduce the well-established properties of the NN interaction in the framework of constituent quarks using the Confined One-Gluon Exchange Potential (COGEP) and Instanton-Induced Interaction (III) potential. As will be shown, the model successfully reproduces the short-range repulsion and intermediate-range attraction. The nature of and the motivation to incorporate the above potentials are also discussed. The primary objective of the present work is to study the interaction between two nucleons (NN interaction) with the Relativistic Harmonic Model (RHM) using the Resonating Group Method (RGM) in Born-Oppenheimer approximation. We analyze NN interaction in the singlet ($^{1}S_{0}$) and triplet ($^{3}S_{1}$) channels and juxtapose the results; the effect of COGEP and III on the adiabatic NN interaction potential are also studied. The results obtained are compared with the experimentally known qualitative features of NN interactions at various ranges. In addition, the contributions of the color magnetic part of COGEP and III to the splitting of the ($^{1}S_{0}$) and ($^{3}S_{1}$) states are studied.

\section{Model and Methodology}
\subsection{The Relativistic Harmonic Model (RHM)}
We use RHM for confinement of quarks \cite{Khadkikar:1985pr} by treating quarks as Dirac fermions confined to Lorentz scalar plus vector potentials. The Dirac equation for a scalar plus vector potential is,
\begin{equation}
[{\bm \alpha}.{\bm p}+\beta (m+S({\bm r}))+V({\bm r})]\psi=E\psi
\end{equation}
where, $S(\bm r)$ is the scalar potential and $V(\bm r)$ is the time component of the vector potential. 

The essential feature of such a potential is that it provides a very sensible framework for mesons as well as baryons. Also, the Klein paradox does not appear for such a potential. The RHM requires both the Lorentz scalar cum vector potentials to be harmonic oscillator potentials. The Lorentz scalar plus vector harmonic oscillator potential is chosen as,
\begin{equation}
\frac{1}{2}(1+\gamma_{0})\alpha^{2}r^{2}+M
\end{equation}
where $\gamma_{0}=\left[
\begin{array}{cc}
1 & 0\\
0 & -1\\
\end{array}
\right]$, $M$ is a constant mass and $\alpha^{2}$ is the confinement strength parameter. The quark wave function ($\psi$) is,
\begin{equation}
\psi=N \left[
\begin{array}{c}
\phi\\
\frac{\bm \sigma.\bm p}{E+M}\phi\\
\end{array}
\right]\label{bispinor}
\end{equation}
where,
\[ N=\sqrt{\frac{2(E+M)}{3E+M}}\]
Here, $E$ is the energy eigenvalue of a single Dirac particle. Eliminating the lower component of $\psi$ in such a way that it satisfies the harmonic oscillator wave equation, we obtain
\begin{equation}
\left[\frac{{\bm p}^{2}}{(E+M)}+\alpha^{2}r^{2}\right]\phi=(E-M)\phi\label{dirac}
\end{equation}
whose eigenvalues are
\begin{equation}
E_{n}^{2}=M^{2}+(2n+1)\Omega_{n}^{2}
\end{equation}
$\Omega_{n}$ ($n\geq 1$) being the energy dependent ``oscillator size parameter'' and $b$ is the oscillator size parameter.

The above-mentioned parameters of the model are chosen so as to obtain meaningful values for the magnetic moments and rms charge radii of the nucleons \cite{Vinodkumar:1992wu}.

\subsection{The Hamiltonian}

By using the RHM for a six-quark system, the Hamiltonian is transformed to act upon $\phi$ in a center of mass (COM) frame. The Hamiltonian then becomes,
\begin{equation}
H=\sum_{i=1}^{6}\frac{\bm {p}_i^{2}}{(E+M)}-K_{\text{CM}}+V_{\text{conf}}+V_{\text{int}}
\end{equation}
Here, $(E+M)/2$ represents the effective dynamic mass of the quark.
The term $K_{\text{CM}}=\frac{P^{2}}{6(E+M)}$ gives kinetic energy for COM motion.
$V_{\text{conf}}$ is the confinement potential. The interaction potential $V_{\text{int}}$ between the quarks is given by $V_{\text{int}}=V_{\text{COGEP}}+V_{\text{III}}$.\par
The interaction potentials considered here are COGEP and III. The central part of COGEP is \cite{VijayaKumar:1993np},
\begin{eqnarray}
V_{\text{COGEP}}(\bm r_{ij})=\frac{\alpha_{s}N^{4}}{4}\sum_{i<j}\bm\lambda_{i}.\bm\lambda_{j}\left[D_{0}(\bm r_{ij})+\frac{1}{(E+M)^{2}}\times \left(4\pi\delta^{3}(\bm r_{ij})-c^{2}D_{0}(\bm r_{ij})\right)\left(1-\frac{2}{3}\bm\sigma_{i}.\bm\sigma_{j}\right)\right]
\end{eqnarray}

The III is given by \cite{Vanamali:2016qwr},
\begin{eqnarray}\label{viii}
V_{\text{III}}(\bm r_{ij})=\frac{-W}{2}\sum_{i<j}\left[\frac{16}{15}+\frac{2}{5}(\bm\lambda_{i}.\bm\lambda_{j})+\frac{1}{10}(\bm\lambda_{i}.\bm\lambda_{j})(\bm\sigma_{i}.\bm\sigma_{j})\right]\delta^{3}(\bm r_{ij})
\end{eqnarray}

In the above expressions, $\bm r_{ij}$ is the separation between the quarks, $\bm\sigma_{i}$ is the well-known Pauli spin matrix of the $i^{th}$ quark, $\bm\lambda_{i}$ are the color (Gell-Mann) matrices, $M$ and $E$ are the quark mass and energy, $\alpha_{s}$ is the quark - gluon coupling constant and $W$ is the III strength parameter.

In what follows, we shall discuss the role of COGEP and III.

\subsubsection{Confined One - Gluon Exchange Potential (COGEP)}
As discussed above, the fact that no hadron (baryon or meson) has been observed as colored states (\textit{i.e.,} quark confinement) and that gluons themselves are color singlets can only seem to imply that gluon color degrees of freedom must also be confined.

In this background, the current confinement model (CCM) was constructed \cite{Vinodkumar:1992wu,Khadkikar:1987kk} to study gluon confinement in the spirit of RHM. Here, the coupled nonlinear term in the Yang-Mills tensor is treated as a color gluon supercurrent - borrowing the analogy from Landau - Ginzburg theory in superconductivity. This leads to an expression for the supercurrent, not unlike the London equation in superconductivity, treating the gluon fields as a quasi-Maxwellian field. Modeled as an effective theory of small fluctuations around a background solution of the classical Yang-Mills theory in the form of a condensate in the background color fields of QCD, a self-induced color current ($j_\mu=\theta_{\mu}^{\nu}A_\nu=m^2A_{\mu}$) adds an equivalent effective mass to the gluons, modifying the gluon propagators \cite{Vinodkumar:1992wu}.

COGEP has a Coulombic part, a color electric part and a color magnetic part. In equation (7) for the central part of COGEP,
\begin{eqnarray}
\frac{\alpha_{s}N^{4}}{4}\sum_{i<j}\bm\lambda_{i}.\bm\lambda_{j}\left[D_{0}(\bm{r}_{ij})+\frac{1}{(E+M)^{2}}(4\pi\delta^{3}(\bm{r}_{ij})-c^{2}D_{0}(\bm{r}_{ij}))\right]
\end{eqnarray}
is the color electric term and
\begin{eqnarray}
- \frac{\alpha_{s}N^{4}}{4}\sum_{i<j}\bm\lambda_{i}.\bm\lambda_{j} \left[\frac{1}{(E+M)^{2}}(4\pi\delta^{3}(\bm{r}_{ij})-c^{2}D_{0}(\bm{r}_{ij}))\times \frac{2}{3}(\bm\sigma_{i}.\bm\sigma_{j})\right]
\end{eqnarray}
is the color magnetic term.
The form of $D_{0}$ is,
\begin{eqnarray}
D_{0}(\bm{r}_{ij}) = \exp \left(-\frac{{r_{ij}^{2} c_{0}^{2}}}{2}\right) \left(\frac{\alpha_{1}}{|\vec{r}_{ij}|}+\alpha_{2}\right)
\end{eqnarray}
where $c_{0}$, $\alpha_{1}$ and $\alpha_{2}$ are parameters that fit $D_{0}$ \cite{VijayaKumar:1993np}.
Because the color magnetic term is spin dependent, it gives rise to the hyperfine splitting between the baryons. And further, since $N$ and $\Delta$ are the only baryons of spin $\frac{1}{2}$ and $\frac{3}{2}$ respectively, the mass difference between these baryons must arise due to the color magnetic interactions.

\subsubsection{Instanton Induced Interaction (III)}
The QCD Lagrangian gives rise to (nearly) massless quarks. In reality, however, quarks have non-zero mass, which arises as a result of their chiral symmetry being broken spontaneously. To this end, instantons provide a meaningful and realistic mechanism for the breaking of chiral symmetry. Since we are studying those quarks with attributes of a constituent mass, instantons become an essential part of the interaction potential and the overall mechanism itself.

For a massless field, the symmetry of the QCD Lagrangian is: $SU(3)_{R} \times SU(3)_{L} \times U(1)_{V} \times U(1)_{A}$. The current corresponding to the $U(1)_{A}$ symmetry is not conserved. The broken $U(1)_{A}$ symmetry is believed to be due to the effects of instanton \cite{Hooft:1976fv,Shuryak:1984nq}. The instantons couple to the quarks, thus giving rise to a constituent mass whose magnitude is of the order of the mass difference between the baryons $N$ and $\Delta$. The III potential mentioned above is the non-relativistic limit of the central part of this instanton field \cite{Oka:1989ud}. Another motivation to introduce instanton physics into QCD is to see if long-range, non-perturbative instanton effects can explain confinement \cite{Belavin:1975fg}.

Also, introducing III, the strength of one-gluon exchange ($\alpha_{s}$) can be reduced. Therefore, from a quantum field theoretical perspective, the instantons are non-perturbative fluctuations of the gluon fields in a QCD vacuum. For these purposes, instantons are taken to be pseudo-particles which come out as the asymptotic solutions to the QCD Lagrangian. Furthermore, they provide a key mechanism for breaking of the chiral symmetry that results in the acquisition of dynamical mass for light quarks \cite{Oka:1989ud,Shuryak:1984nq}. Therefore, one can account for the dynamical mass of quarks due to Instanton Induced Interaction (III) and hence, III becomes a quintessential part of NN interaction in our study \cite{Oka:1989ud,Vanamali:2016qwr}.

\begin{figure}[h]
	\centering
	\subfloat{\includegraphics[height=2.5in,width=3in]{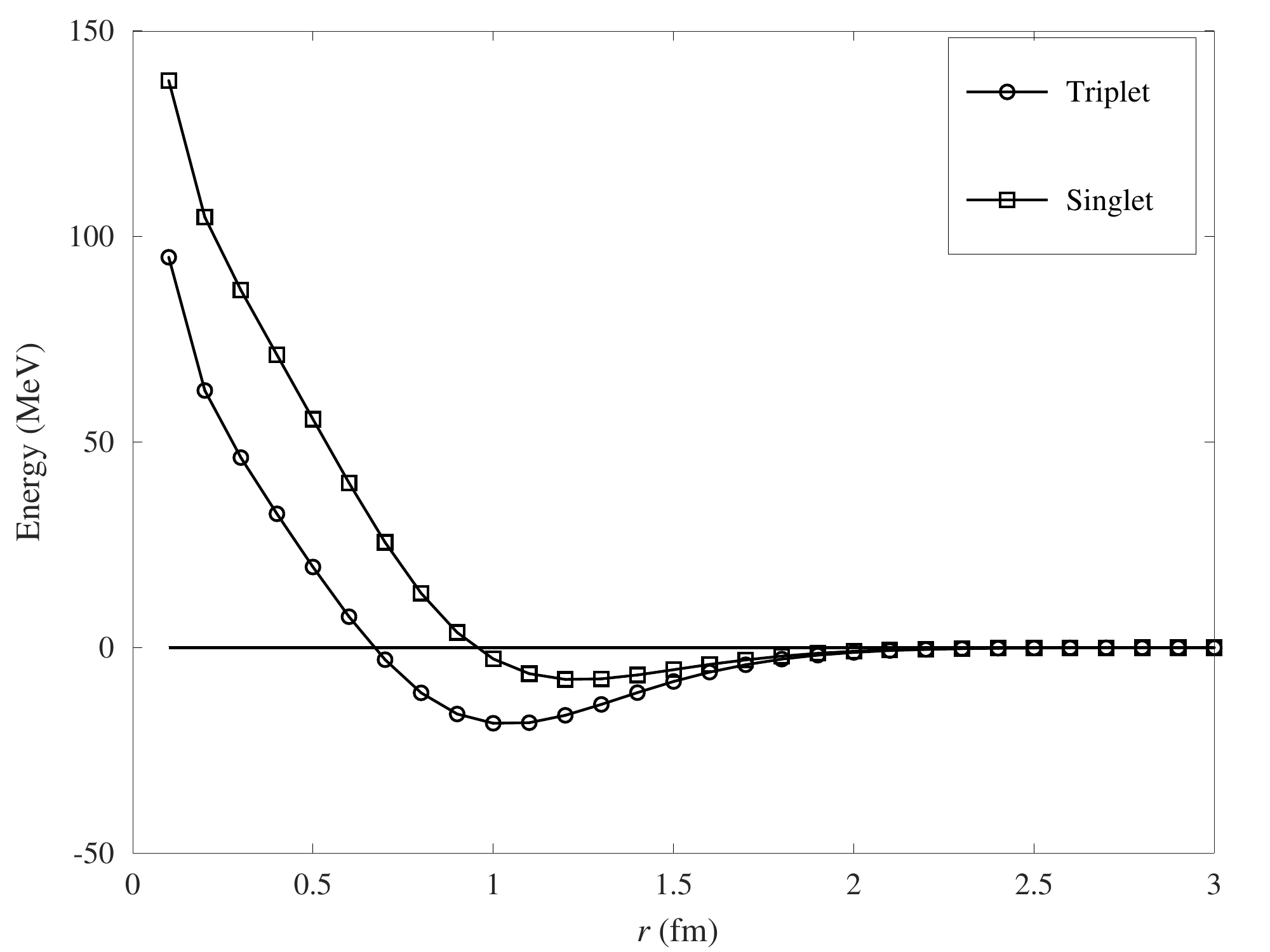}}
	\subfloat{\includegraphics[height=2.5in,width=3in]{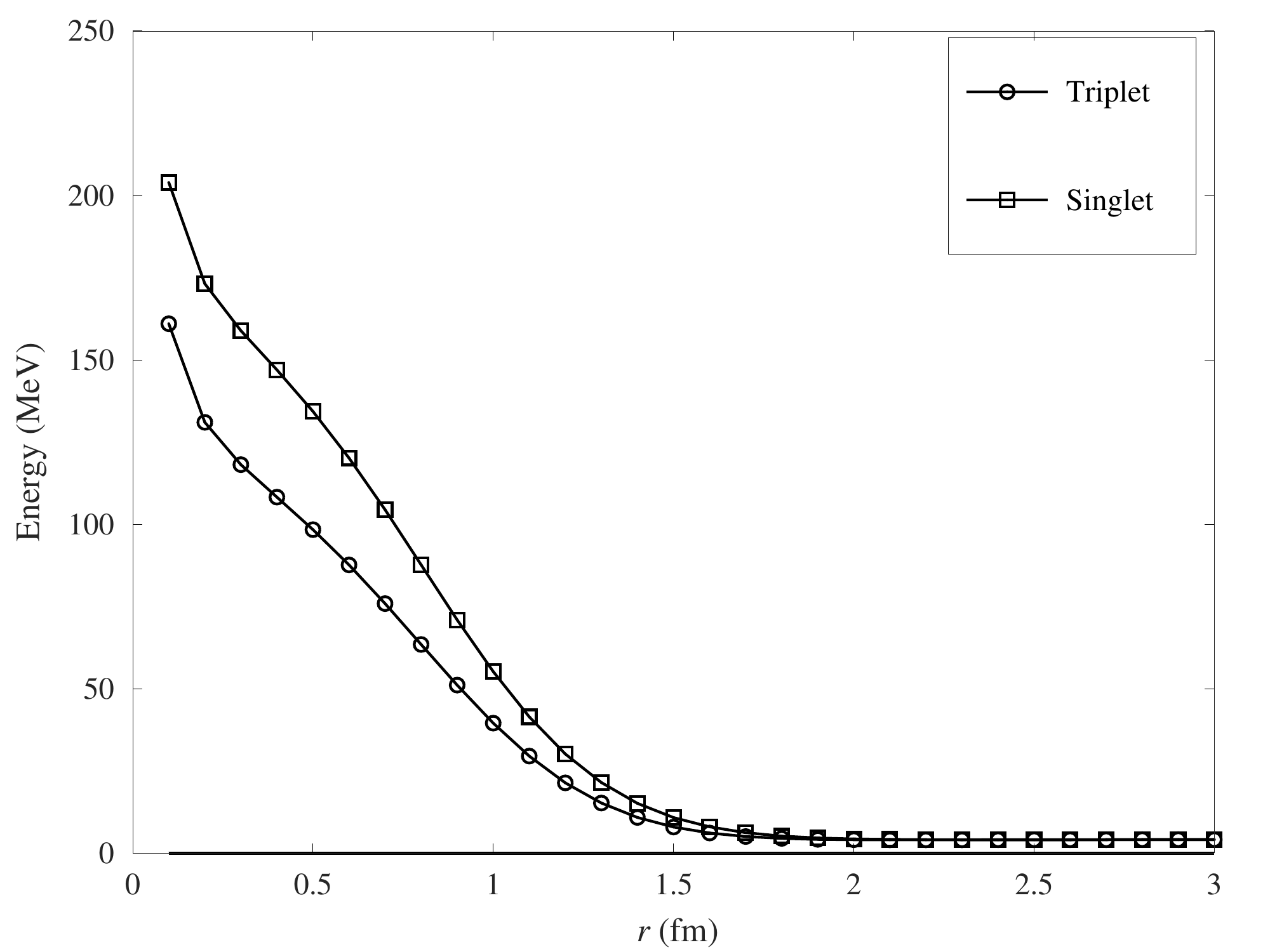}}
	\caption{Plots of the NN potential obtained in the present work (left) and the exchange component of the NN potential (right).}\label{nnpot}
\end{figure}

The III can be derived from the effective instanton Lagrangian that leads to a contact interaction between the constituent quarks in the leading order. The potential derived from the interaction is given by \cite{Oka:1989ud},
\begin{equation}
	V_{\text{III}}(\bm r_{ij}) = -\sum_{i<j}\frac{1}{2}W_{ij}\left(1-P_{ij} \right)\left( 1-\frac{1}{5}(\bm\sigma_i\cdot\bm\sigma_j) \right)\delta(\bm r_{ij})
\end{equation}
where, $P_{ij}$ represents the permutation of the $i^{th}$ and $j^{th}$ quarks and the $W_{ij}$ is the strength of the corresponding interaction. Representing the permutation operator in terms of $\bm\sigma_i$ and $\bm\lambda_i$, we get eq. (\ref{viii}).

\subsection{Resonating Group Method}

The resonating group method (RGM) is used to calculate the potential for the NN interaction. The expression for the interaction potential is given by,
\begin{equation}
	E=\frac{\langle\psi|HA|\psi\rangle_{l=0}}{\langle\psi|A|\psi\rangle_{l=0}}
\end{equation}
where, H is the Hamiltonian, $\psi$ is the wave-function of the two-nucleon system and A is the anti-symmetrization operator. The anti-symmetrization operator guarantees that the total wave-function of the NN system is anti-symmetric and is given by \cite{Oka:1989ud}:,
\begin{equation}
	A=\frac{1}{10}(1-9P_{36}^{OSTC})
\end{equation}
where, $P_{36}^{OSTC}$ is the permutation operator for quarks 3 and 6 and OSTC stands for orbital, spin, isospin and color respectively. Thus, $P_{36}^{\text{OSTC}}$ operator exchanges the orbital, spin, isospin and color quantum numbers of the quarks 3 and 6 \cite{VijayaKumar:1993np}. The anti-symmetrization operator separates the contributions to the interaction potential in to two parts - direct and exchange. The direct part represents the internal dynamics of each nucleon and exists at all separations. The exchange part represents the effects overlapping of the nucleon wave functions and is present only in region of interaction. The interaction potential is obtained by removing asymptotic contributions using the adiabatic approximation.

We choose the Harmonic oscillator wave function as the trial wave function for the orbital part of the calculations:
\begin{equation}
\phi(\bm r_{i},\bm S_{I}) = \frac{1}{(\pi b^{2})^{3/4}} \exp\left(-\frac{1}{2b^{2}}\left(\bm r_{i} - \frac{\bm S_{I}}{2}\right)^{2}\right)
\end{equation}
where, $\bf{S}_I$ is the relative separation between the nucleons. The total wave function is given by,

\begin{equation}
	\psi = \prod_{i=1}^3 \phi(\bm r_{i},\bm S_{I}) \phi(\bm r_{i+3},-\bm S_{I})
\end{equation}

\section{Results and Discussion}

\begin{figure}[h]
	\centering
	\subfloat{\includegraphics[height=2.5in,width=3in]{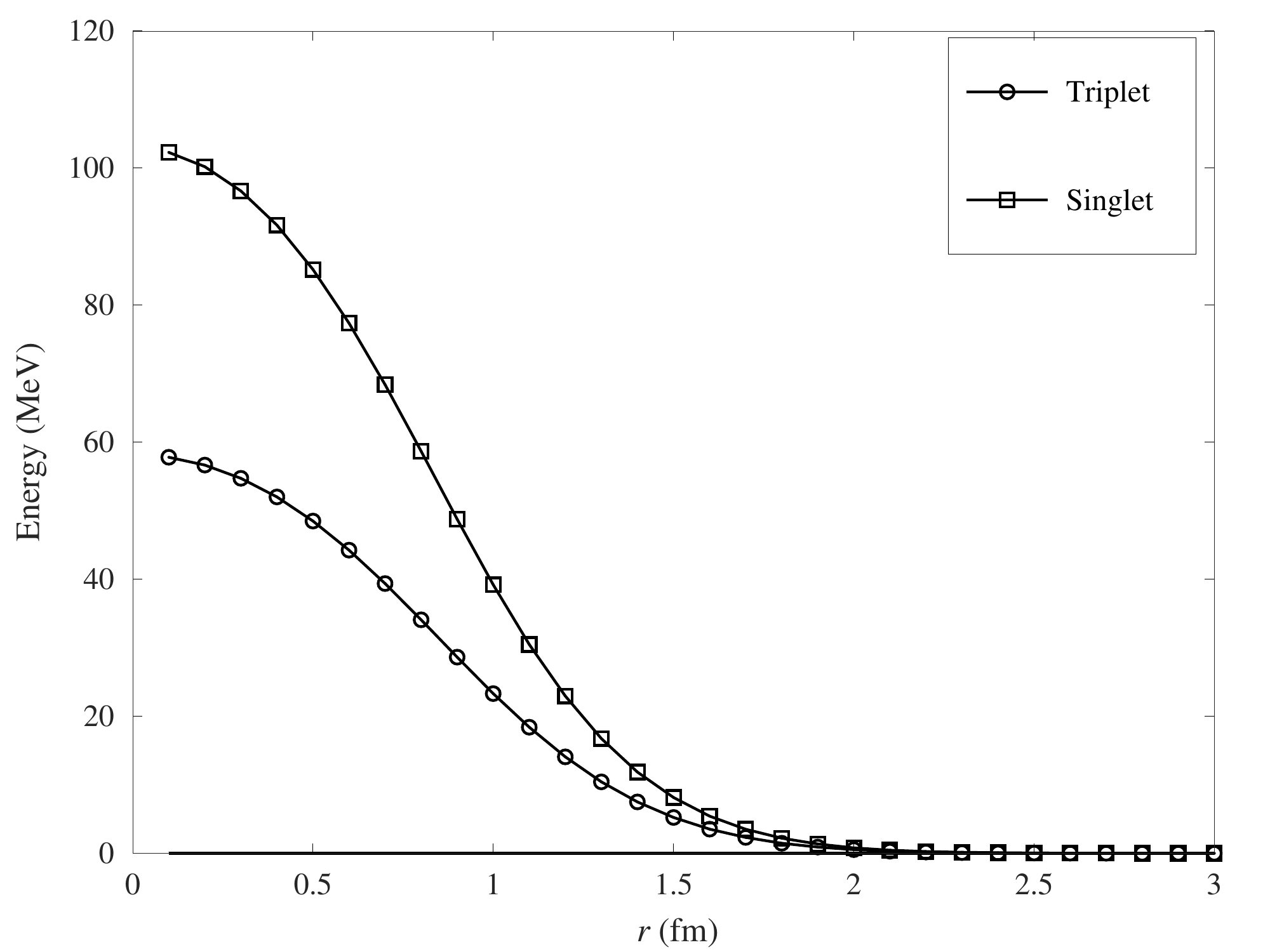}}
	\subfloat{\includegraphics[height=2.5in,width=3in]{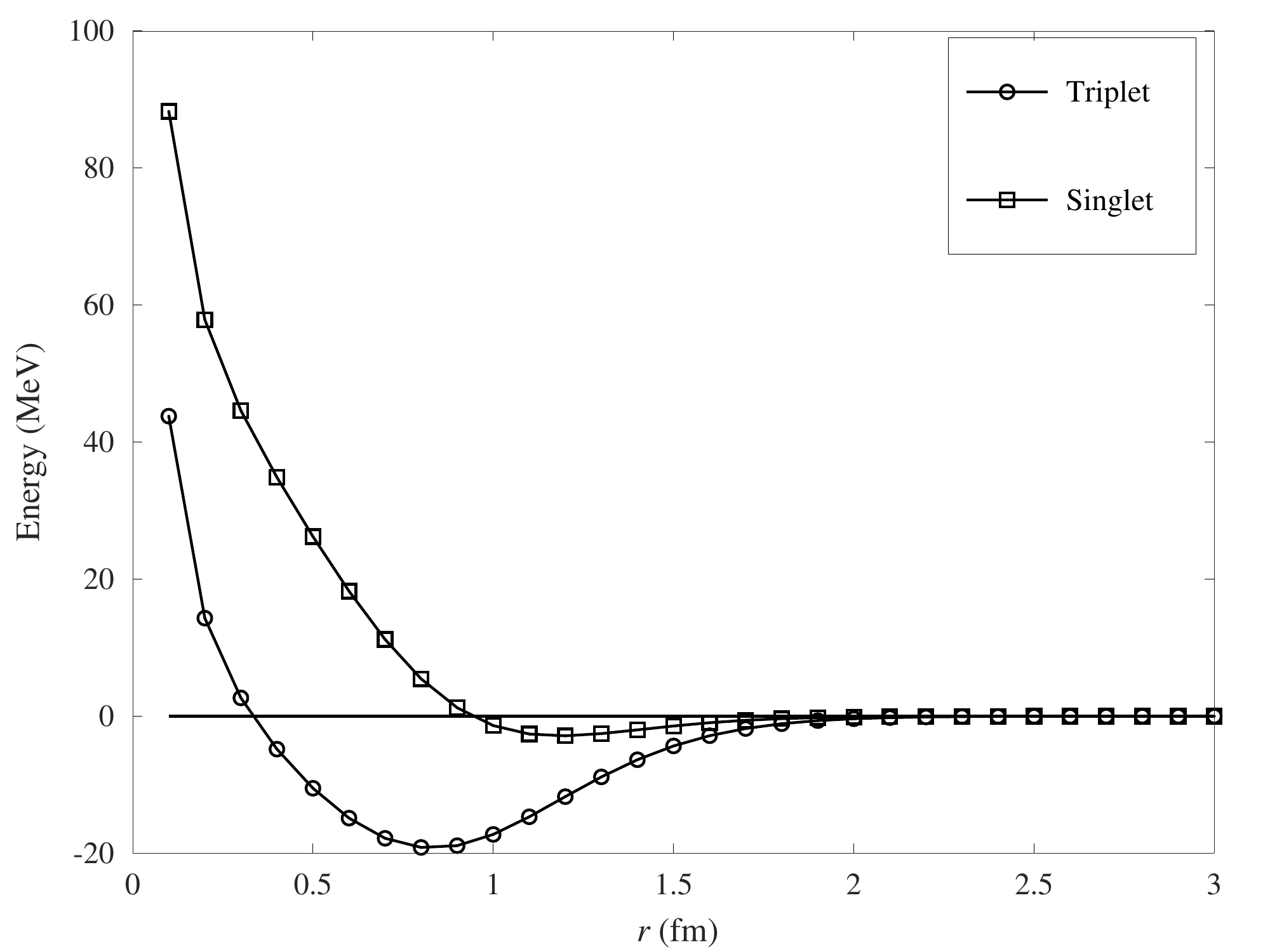}}
	\caption{Plots of the contribution of color magnetic exchange part COGEP (left) and the total contribution of COGEP (right) to NN potential.}\label{cmag}
\end{figure}

\begin{figure}[h]
	\centering
	\subfloat{\includegraphics[height=2.5in,width=3in]{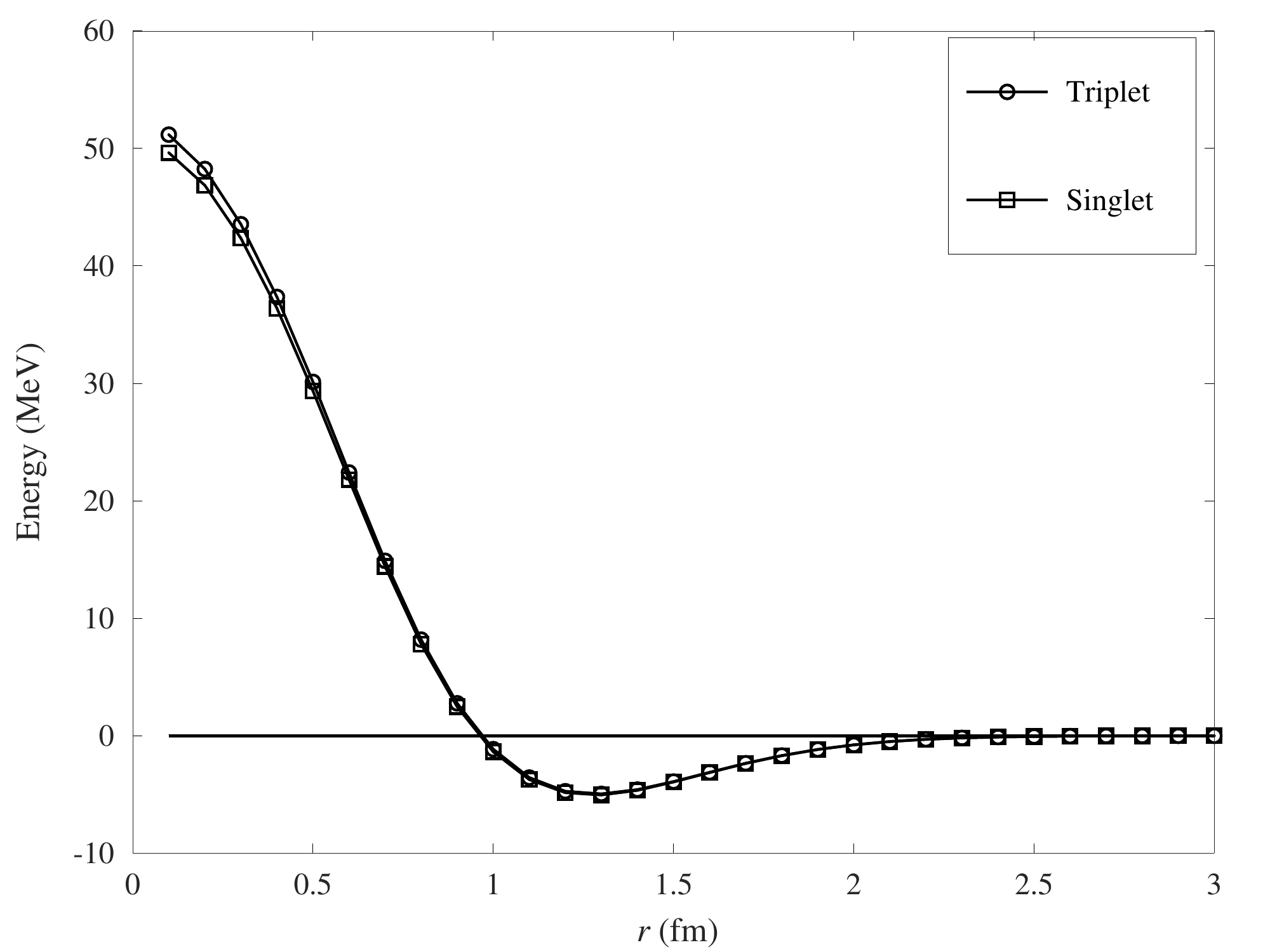}}
	\subfloat{\includegraphics[height=2.5in,width=3in]{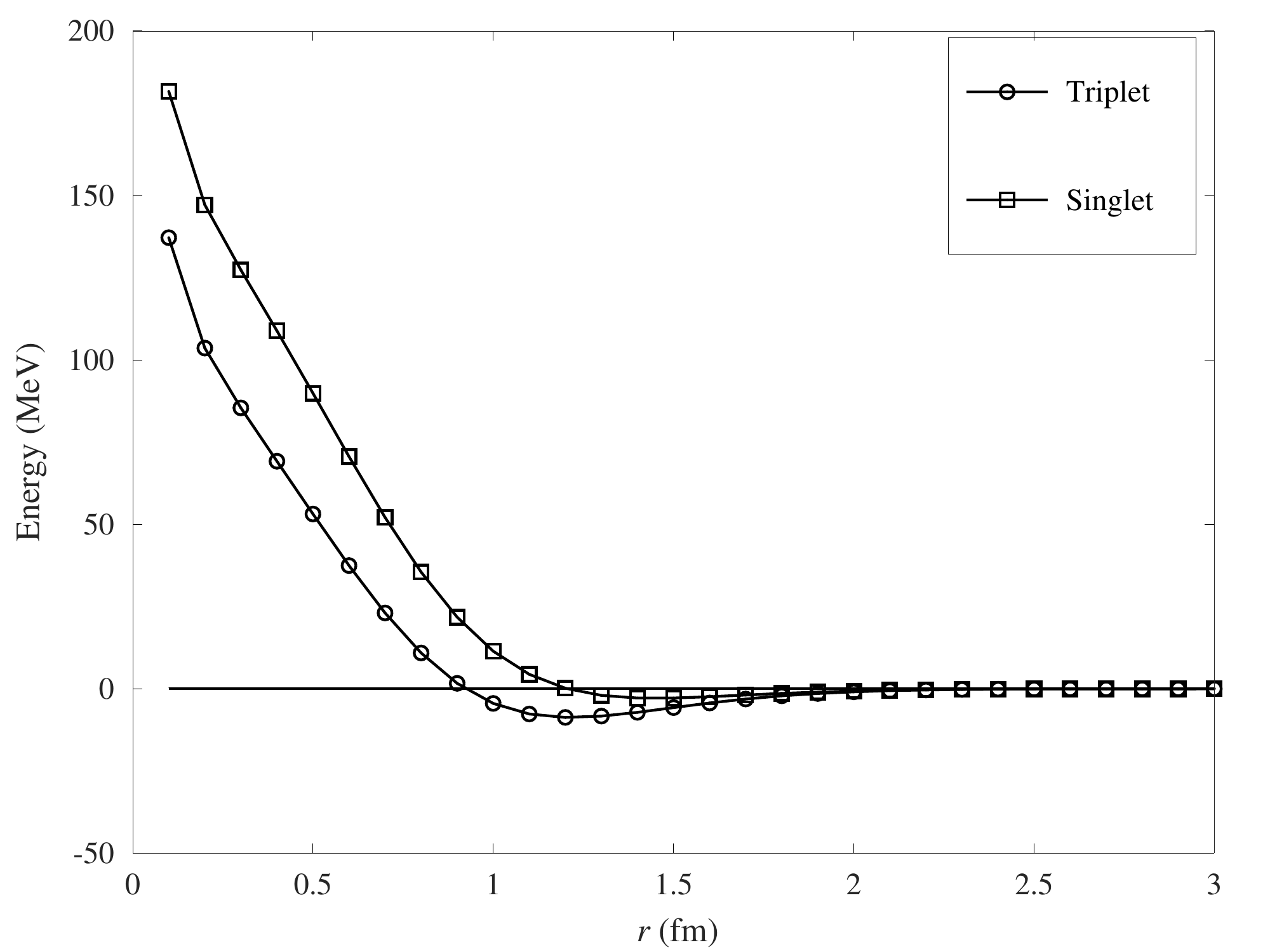}}
	\caption{Plots of the NN potential without COGEP (left) and without III (right).}\label{nocogep}
\end{figure}

\begin{table*}[h]
\centering
\begin{tabular}{c|c}
	Parameter & Value \\
	\hline\\
	$M$ & $160.6$ MeV\\
	$E$ & $ 40.0$ MeV\\
	$\alpha^{2}$ & $ 200.01 $ MeV\\
	$b$ & $0.6$ fm\\
	$\alpha_{s}$ & $ 6.5 $\\
	$c$ & $0.3$ fm$^{-1}$\\
	$\alpha_1$ & $1.035994$\\
	$\alpha_2$ & $2.016150$ MeV\\
	$c_0$ & $1.7324$ fm$^{-1}$\\
	$W$ & $67.67$ MeV fm$^{3}$
\end{tabular}
	\caption{Parameters used in the model.}\label{table}
	\end{table*}

Our model has ten parameters  - the quark mass and energy eigen value ($M$ and $E$), confinement strength ($\alpha^{2}$), the harmonic oscillator size parameter ($b$), the quark-gluon coupling constant ($\alpha_{s}$), III strength parameter ($W$), CCM parameter ($c$), $c_{0}$, $\alpha_{1}$ and $\alpha_{2}$ - the parameters that characterize the function $D_{0}(r)$ \cite{Khadkikar:1991fv,VijayaKumar:1993np}. The parameters and their values are listed in the table \ref{table}. The oscillator size parameter was chosen to be $0.6$ fm and is consistent with the experimental results \cite{Shimizu:1989ye}. The chiral symmetry breaking provides a dynamical mass of $\sim 300\text{ MeV}$ to the constituent quarks. As evident from Eq. (\ref{bispinor}) and Eq. (\ref{dirac}), the dynamical mass of the constituent quarks is given by the sum of their energy eigen value in the confining harmonic potential and the mass parameter ($M$). The values of the parameters $E$ and $M$ are fitted from hadron spectroscopy. $\alpha^{2}$ is fixed from the stability condition for the nucleon mass as a function of the size parameter $b$  \cite{Vinodkumar:1992wu}.\par

The total NN potential obtained from the present study is shown in fig. (\ref{nnpot}). The plot shows a large repulsion arising in the short range and an attraction in the intermediate range. The exchange part of the NN potential is completely repulsive in both the $^{1}S_{0}$ and $^{3}S_{1}$ channels. The dominant contributions to the NN interaction potential comes from the kinetic energies of the quarks, color magnetic interactions, and quark confinement. The color magnetic part of COGEP gives short-range repulsion in the $^{1}S_{0}$ and $^{3}S_{1}$ channels as shown in fig. (\ref{cmag}). The exchange kernels of $\delta^{3} (\bm r_{ij})$ dominate over the exchange kernels of $c^{2}D_{0} (\bm r_{ij})$ at short-range distances, thus producing short-range repulsion \cite{Shimizu:1989ye}; whereas, the exchange kernels of $c^{2}D_{0} (\bm r_{ij})$ dominate over that of $\delta^{3} (\bm r_{ij})$ at intermediate and long ranges. It is also observed that there's no significant contribution from the color electric terms. The color electric term consists of a radial matrix element, which is the same for the two hadron $(2(0s)^3)$ state and for the six - quark $((0s)^6)$ state. The difference arises from the color interactions $\bm\lambda_{i}.\bm\lambda_{j}$.  But, the expectation value of the $\bm\lambda_{i}.\bm\lambda_{j}$ depends only on the number of quarks. Hence, the color electric elements of COGEP, III and the confinement term do not contribute NN adiabatic potential. The decisive contribution of COGEP can be seen when the total NN potential is plotted without COGEP as shown in fig. (\ref{nocogep}). It can be seen that COGEP provides a repulsion of about 150 MeV in the short range, which is significantly reduced to just about 50 MeV without COGEP. This repulsion is larger than the repulsion provided by the conventional one gluon exchange potential \cite{Vanamali:2016qwr}. Also, COGEP provides an attraction of about 20 MeV in the intermediate range. Thus, the intermediate attraction must arise from the direct components of the potential.\par

 As seen in a previous study, the direct part of the III provides a large state independent attraction \cite{Vanamali:2016qwr}. The direct part of the color magnetic term shows state-independent attraction at short range. The exchange part shows repulsion. It can be seen that III provides a net attraction predominantly, up to $\approx$2 fm. Therefore, III is an attractive potential. A plot of the total potential without III shows a surplus in the short-range repulsion of about 50 MeV and the intermediate-range attraction is reduced by about 20 MeV. Further, the attractive minimum is situated at about 1 fm with III, which is slightly shifted to about 1.3 fm without it. This reaffirms the attractive nature of III. As for the potentials, short-range repulsion arises from the kinetic energy and the exchange part of the color magnetic terms of the potential. \par

Summarizing, we have studied the NN interaction using a relativistic harmonic model that includes the confinement of the gluons. The adiabatic NN potential for the singlet and triplet states have been obtained in the Born - Oppenheimer approximation. We observe that the Hamiltonian is highly non-local. The gluon confinement is shown to increase the short range repulsion. The III is shown to be essential for the intermediate range attraction.

\end{document}